\documentclass[notitlepage]{revtex4-1} 
\usepackage{graphicx}

\usepackage{hyperref} 					 
\usepackage{listings} 					 
\usepackage{color}    					 
\usepackage{natbib}   					 
\usepackage{subfig} 					 

\usepackage{booktabs}
\usepackage[ansinew]{inputenc}				     
\usepackage{psfrag}
\usepackage{pstricks}
\usepackage{amsfonts}
\usepackage{amsmath}
\usepackage{bm}
\usepackage{lipsum}
\usepackage{pgf}
\usepackage{tikz}
\usetikzlibrary{patterns}

\usepackage{array}
\newcolumntype{C}[1]{>{\centering\arraybackslash}m{#1}}

\usepackage{algpseudocode}
\usepackage{color}

\usepackage{filecontents}

\usepackage{algpseudocode}
\usepackage{color}
\usepackage{soul}

\begin{document}
	
	\title{Fullwave Maxwell inverse design of axisymmetric, tunable, and multi-scale multi-wavelength metalenses}
	
	\author{Rasmus E. Christiansen,$^{1,2,3}$}
	\email{raelch@mek.dtu.dk}
	
	\author{Zin~Lin,$^{3}$}  \author{Charles~Roques-Carmes,$^{4}$} \author{Yannick~Salamin,$^{4}$} \author{Steven~E.~Kooi,$^{6}$} \author{John~D.~Joannopoulos,$^{5,6}$} \author{Marin~Solja\v{c}i\'{c},$^{4,5}$} \author{Steven~G.~Johnson$^{3,4}$}
	
	\affiliation{$^{1}$ Department of Mechanical Engineering, Technical University of Denmark, Nils Koppels All\'{e}, Building 404, 2800 Kongens Lyngby, Denmark}
	\affiliation{$^{2}$ NanoPhoton---Center for Nanophotonics, Technical University of Denmark, {\O}rsteds Plads 345A, DK-2800 Kgs. Lyngby, Denmark.}
	\affiliation{$^{3}$ Department of Mathematics, Massachusetts Institute of Technology, Cambridge, MA 02139,
		USA}
	\affiliation{$^{4}$ Research Laboratory of Electronics, Massachusetts Institute of Technology, 50 Vassar St., Cambridge, MA}
	\affiliation{$^{5}$ Department of Physics, Massachusetts Institute of Technology, Cambridge, MA 02139, USA}
	\affiliation{$^{6}$ Institute for Soldier Nanotechnologies, 500 Technology Square, Cambridge, MA}
	
	
	
	
	\begin{abstract}
		We demonstrate new axisymmetric inverse-design techniques that can solve problems radically different from traditional lenses, including \emph{reconfigurable} lenses (that shift a multi-frequency focal spot in response to refractive-index changes) and  {\emph{widely separated}} multi-wavelength lenses ($\lambda = 1\,\mu$m and $10\,\mu$m). We also present experimental validation for an axisymmetric inverse-designed monochrome lens in the near-infrared fabricated via two-photon polymerization. Axisymmetry allows fullwave Maxwell solvers to be scaled up to structures hundreds or even thousands of wavelengths in diameter before requiring domain-decomposition approximations, while multilayer topology optimization with $\sim 10^5$ degrees of freedom can tackle challenging design problems even when restricted to axisymmetric structures.
	\end{abstract}
	
	\maketitle
	
	\section{Introduction}
\begin{figure}[h!]
		\centering\includegraphics[width=1.0\linewidth]{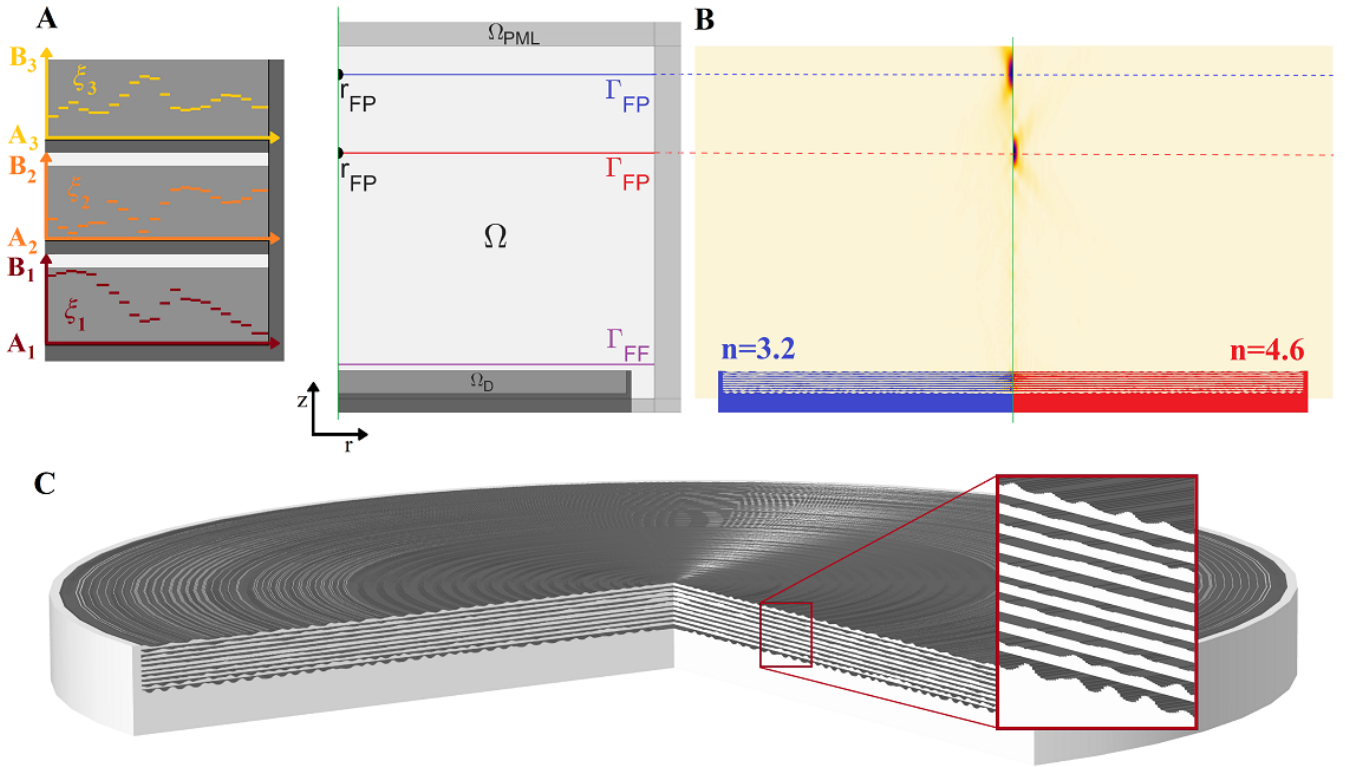}
		\caption{\textbf{A)} Sketch of the multilayer design domain [left] and model domain [right]. \textbf{B)} Illustration of an active metalens, operating at $\lambda=10 \ \mu$m at $n=3.2$ [left] and $n=4.6 + 0.01\mathrm{i}$ [right], showing the max-normalized transmitted $\vert \textbf{E} \vert^2$-field (thermal colormap) in the (x,y)-plane through the center of the lens. \textbf{C)} 3D rendering of the active metalens geometry. \label{FIG:DESIGN_PROBLEM_ILLUSTRATION}}
	\end{figure}

	In this paper, we demonstrate that axisymmetric metalenses can be designed with fullwave Maxwell simulations (as opposed to the scalar-diffraction~\cite{meem2020large} or domain-decomposition approximations~\cite{pestourie2018inverse} used in prior metasurface designs), for $> 100$~wavelengths ($\lambda$) in diameters, combined with multilayer variable-height topology optimization (TO) with $\gtrsim 10^4$ degrees of freedom ($\sim 10$ per $\lambda$ per layer) as shown in Fig.~\ref{FIG:DESIGN_PROBLEM_ILLUSTRATION}. The capability and flexibility of our design approach is demonstrated by solving two challenging new design problems with 10-layer metalenses. First (in Sec.~\ref{SEC:LENS_OOMDW}), we design a \emph{multi-scale} metalens that \emph{simultaneously} focuses $\lambda = 1\,\mu$m and $\lambda=10\,\mu$m at the same diffraction-limited focal point (numerical aperature NA $=0.85$, Strehl ratios of $0.60$ and $0.84$ and efficiencies of $82$\% and $95$\%, respectively). Second (in Sec.~\ref{SEC:LENS_AM}), we design an \emph{active} metalens that shifts its achromatic multi-wavelength (three $\lambda$s over a 6\% bandwidth) focal spot from NA $=0.7$ to $0.8$ as the index of the material (GSS4T1~\cite{ZHANG_ET_AL_2019}) is changed from $n=3.2$ to $n=4.6$ (thermally or optically), in contrast to previous work that showed only monochromatic reconfigurability~\cite{shalaginov2019reconfigurable}. As a proof of concept, we also show (in Sec.~\ref{SEC:SINGLE_LAYER_EXPERIMENTAL_VALIDATION}) an experimental realization of single-layer axisymmetric TO-designed metalens for $\lambda = 1550$~nm, fabricated by two-photon polymerization 3D-printing (Nanoscribe Professional GT), demonstrating that such variable-height surfaces are manufacturable. As discussed in Sec.~\ref{SEC:CONCLUSION}, our approach could easily be scaled to much larger diameters and number of layers, and the vast number of design degrees of freedom coupled with the lack of approximations makes it a uniquely attractive method for the most difficult metasurface inverse designs.
	
	Flat-optics metalenses have received widespread attention due to their potential for achieving multiple functionalities within an ultra-compact form factor~\cite{yu2014flat,khorasaninejad2016metalenses,khorasaninejad2017metalenses,aieta2015multiwavelength,chen2018broadband}. Prior work on metalens design has largely focused on exploiting local resonant conditions~\cite{yu2014flat} under locally periodic approximation (LPA), using rather small unit cells ($\lesssim \lambda$)~\cite{pestourie2018inverse,lin2019topology}. Recently, it has been shown that the unit-cell approach with LPA can lead to fundamental limitations on metalens performance~\cite{chung2020high,presutti2020focusing}, whereas some of these limitations may be mitigated by using overlapping boundaries, perfectly matched layers or larger domains~\cite{lin2019overlapping,phan2019high} some may not. Meanwhile, axisymmetric multi-level diffractive lenses (MDL) have been proposed as an alternative to metalenses for achieving enhanced functionalities~\cite{BANERJI_ET_AL_2019,meem2020imaging}; however, MDL designs utilize scalar diffraction theory subject to locally uniform approximation, neglecting multiple scatterings or resonant phenomena, and, thus, have limited design complexity and physical behavior~\cite{engelberg2020advantages}. In contrast to previous works, our approach considers \emph{axisymmetric multilayered freeform metalenses} which can be modelled by rigorous fullwave Maxwell equations without any uncontrolled approximations.
	
	The prospect of fabricating single-layer metasurfaces with traditional single-step lithography processes is very promising for large-scale high-throughput integration \cite{khorasaninejad2016metalenses, khorasaninejad2017metalenses}. However, single-layer metasurfaces have been limited in their functionality to narrow angular and spectral bandwidths of operation, with some progress towards chromatic \cite{chen2018broadband} and geometrical aberration correction \cite{groever2017meta}. Achieving truly multifunctional metasurfaces requires more advanced designs, such as closely-packed multilayer structures~\cite{lin2018topology,lin2019topology,lin2019overlapping,chung2020high}. Recently, there has been a surge in interest in fabricating multilayer metasurfaces \cite{mansouree2020multifunctional, zhou2018multilayer}, bolstered by advances in inverse-designed nanophotonics. However, these designs can only be fabricated with more advanced fabrication techniques, such as multi-step lithography, or multi-photon lithography. Multi-photon lithography/polymerization is a technique that enables the fabrication of sub-micron (down to $\sim$ 150~nm) arbitrary three-dimensional structures. Two-photon polymerization enabled the demonstration of three-dimensional chiral/helical structures \cite{gansel2009gold, wu20193d} and was more recently applied to the fabrication of supercritical lenses \cite{fang2020multilevel, coompson20183d} and to the demonstration of full three dimensional control of optical fields using a metasurface \cite{ZHAN_ET_AL_2019}. Nonetheless, the possibility of 3D printing inverse-designed metasurfaces with two-photon lithography processes remains largely unexplored. In this work, we realize a proof-of-concept experiment with an inverse-designed, freeform, single-layer metalens working at $\lambda = 1550$ nm fabricated via two-photon polymerization.
	
	The use of TO as a tool for inverse design in nanophotonics has increased steadily over the last two decades~\cite{JENSEN_SIGMUND_2011,MOLESKY_2018} with a recent application to metalens design~\cite{lin2018topology}. Our proposed multilayer metalens design framework utilizes density-based topology optimization~\cite{BOOK_TOPOPT_BENDSOE} as the inverse design tool. Rather than allowing fully free-form 3D designs, not amenable to nano-scale fabrication, we propose using TO to control the radial height-variation of the $\mathcal{N}$-layers constituting the lens. In addition, we propose using a filtering technique~\cite{BOURDIN_2001,LAZAROV_2011} combined with a threshold operation to limit the gradient of the height variations, making it possible to ensure that they comply with fabrication constraints. 
	
	\section{Model and design problem} \label{SEC:MODEL_PROBLEM}
	
	The design problem is modelled in an axisymmetric domain, $\Omega$, sketched in Fig.~\ref{FIG:DESIGN_PROBLEM_ILLUSTRATION}A(right). The model domain contains a designable region, $\Omega_D$ (gray), where the metalens is placed on a solid material surface (dark gray). The sketch also indicates the plane, $\Gamma_{\mathrm{FF}}$ (magenta line), used for computing the far-field transformation in Eq.~\ref{EQN:FAR_FIELD_COMPUTATION}, the focal plane(s), $\Gamma_{\mathrm{FP}}$ (blue and red lines), and the focal spot(s), $\textbf{r}_{\mathrm{FP}}$ (black dots), of the lens. Finally, the model domain is truncated using a perfectly matching layer in $\Omega_{\mathrm{PML}}$ \cite{BERENGER_1994,CHEW_1997}. The lens-design itself consists of $\mathcal{N}$ layers of material (Fig.~\ref{FIG:DESIGN_PROBLEM_ILLUSTRATION}A(left)), each with a variable height controlled by the design field, $\bar{\xi}_L(r)$. Each designable layer is separated from the next by a layer of air (light gray) and a layer of solid material (dark gray) of fixed thicknesses. 
	
	The physics is modelled in $\Omega$ using the Maxwell Equations~\cite{BOOK_EM_GRIFFITHS} assuming time-harmonic behavior. Doing so, we capture the full-wave behavior of the electromagnetic field without simplifying assumptions, thus enabling the exploitation of the full-wave behavior to design metalenses exerting precise control of the electromagnetic field. To make the model problem numerically tractable for large design domains, we assume an axisymmetric geometry. This enables the reduction of the full 3D Maxwell Equations to their 2D axisymmetric counterparts~\cite{BOOK_FEM_EM_JIN}, therefore significantly reducing the computational effort required to solve the model problem at the cost of a geometric restriction on the design. 
	
	The model problem is solved using the scattered-field formulation, $\textbf{E} = \textbf{E}_\mathrm{b} + \textbf{E}_\mathrm{s}$, where the background field, $\textbf{E}_{\mathrm{b}}$, is taken to be a planewave propagating along the z-axis (decomposed into two counter-rotating circularly polarized waves). The model problem is discretized using the Finite Element Method (FEM)~\cite{BOOK_FEM_EM_JIN} and solved using COMSOL Multiphysics v5.5 \cite{COMSOL55}.
	
	A far-field transformation~\cite{taflove2005computational} may be used to compute the electric field at any point in space given knowledge of the field in the plane $\Gamma_{\mathrm{FF}}$ above the lens, see Fig. \ref{FIG:DESIGN_PROBLEM_ILLUSTRATION}. The use of this transformation removes the need for simulating the spatial domain between the lens and focal point, hereby significantly reducing computational cost. The far field transformation may be written as,
	
	\begin{eqnarray} \label{EQN:FAR_FIELD_COMPUTATION}
	\textbf{E}_\mathrm{FF}(\textbf{r}) = \int_{\Gamma_{\mathrm{FF}}} \textbf{G}^\textbf{E}(\textbf{r},\textbf{r}') \textbf{K}(\textbf{r}') +\textbf{G}^\textbf{H}(\textbf{r},\textbf{r}') \textbf{J}(\textbf{r}')  \ \mathrm{d}\textbf{r}'. \label{EQN:FAR_FIELD} 
	\end{eqnarray}
	
	Here $\textbf{E}_\mathrm{FF}(\textbf{r})$ denotes the electric far field at the point $\textbf{r}$, $\textbf{G}^\textbf{E}(\textbf{r},\textbf{r}')$ and $\textbf{G}^\textbf{H}(\textbf{r},\textbf{r}')$ denotes the electric and magnetic field Green's functions, respectively. Finally $\textbf{K}(\textbf{r}') $ and $\textbf{J}(\textbf{r}') $ denote the equivalent magnetic and electric surface currents computed from the electric and magnetic near field obtained by solving the model problem.
	
	The figure of merit (FOM) used in the design process is the electric field-intensity at the focal point, $\textbf{r}_{\mathrm{FP}}$. The design problem is formulated as the following continuous constrained optimization problem,
	\begin{eqnarray} \label{EQN:DESIGN}
	&\underset{\xi(\textbf{r}) \in [0,1]}{\max} \ \ \ &\Phi(\xi) =  \vert \textbf{E}_{\mathrm{FF}}(\textbf{r}_{\mathrm{FP}},\xi) \vert^2, \label{EQN:OBJECTIVE_FUNCTION} \\ \label{EQN:DESIGN_CON_1}
	&\mathrm{s.t.} \ \ \ &A_L \leq \xi_L(r) \leq B_L, \ \ L \in \lbrace 1,2,...,\mathcal{N} \rbrace, \ \ \mathcal{N} \in \mathbb{N}
	\end{eqnarray}
	
	Here $\xi_L(r)$ denotes a radially-varying design field, which controls the thickness of the $L$'th layer of the metalens. The electric field at the focal point, $\textbf{E}_{\mathrm{FF}}(\textbf{r}_{\mathrm{FP}},\xi)$, is computed using the solution to the physical model problem and Eq.~\ref{EQN:FAR_FIELD_COMPUTATION}. 
	
	We propose using a standard PDE-filter~\cite{LAZAROV_2011} to limit the layer-thickness gradient, by applying it to $\xi_L(r)$ through the choice of filter radius, $r_f$. After filtering we propose using $\xi_L(r)$ to control the layer height through the smoothed threshold operation \cite{WANG_ET_AL_2011} as,
	\begin{eqnarray} \label{EQN:HEAVISIDE}
	\bar{\xi}_L = 1 - \frac{\tanh(\beta \cdot \xi_L) + \tanh(\beta \cdot (z_L - \xi_L))}{\tanh(\beta \cdot \xi_L) + \tanh(\beta \cdot (B_L - \xi_L))}, \ \ \beta \in [1,\infty[, \ \ \xi_L \in [A_L,B_L], \ \ z_L \in [A_L,B_L].  \label{EQN:THRESHOLD_FUNCTION}
	\end{eqnarray}
	
	Here $z_L$ denotes the spatial position inside each designable layer. The value $z_L = A_L$ corresponds to the bottom of the designable region in the $L$'th layer, and the value $z_L = B_L$ corresponds to the top of the designable region in the $L$'th layer. The threshold sharpness is controlled by $\beta$. In the limit of $\beta \rightarrow \infty$ the field $\bar{\xi}_L$ takes the value 0 when $z_L > \xi_L$ and the value 1 when $z_L < \xi_L$. A continuation approach may be used to gradually increase $\beta$ during the inverse design process to enforce a 0/1 final design.
	
	The field $\bar{\xi}_L$ is used to interpolate the relative permittivity, $\varepsilon_r(\textbf{r})$, in space between the background material and the material constituting the metalens using a linear scheme,
	\begin{eqnarray} 
	\varepsilon_r(\textbf{r}) = \varepsilon_\text{r,bg} + \bar{\xi}_L (\varepsilon_\text{r,lens} - \varepsilon_\text{r,bg}). \label{EQN:INTERPOLATION}
	\end{eqnarray}
	
	\noindent Here $\varepsilon_\text{r,bg}$ (resp. $\varepsilon_\text{r,lens}$) denotes the relative permittivity of the background (resp. lens).   
	
	The design problem, Eqs.~(\ref{EQN:OBJECTIVE_FUNCTION}-\ref{EQN:DESIGN_CON_1}), is solved using the Method of Moving Asymptotes~\cite{SVANBERG_2002}, for which the sensitivites of the FOM are computed using adjoint sensitivity analysis~\cite{TORTORELLI_ET_AL_1994}. Details regarding the modelling and optimization process as well as the parameter choices for each example are found in Appendix~A and Appendix~B, respectively.
	
	The final designs are all evaluated numerically using a high resolution model by exciting the lens using a linearly polarized planewave decomposed into two counter-rotating circularly polarized waves introduced in the model using a first order scattering boundary condition. An example of a reconfigurable metalens operating at $\lambda = 10~\mu$m for two different refractive indices is shown in Fig.~\ref{FIG:DESIGN_PROBLEM_ILLUSTRATION}B.
	
	\section{Multi-scale multi-wavelength multilayer metalens} \label{SEC:LENS_OOMDW}
	
\begin{figure}[h!]
		\centering\includegraphics[width=1.0\linewidth]{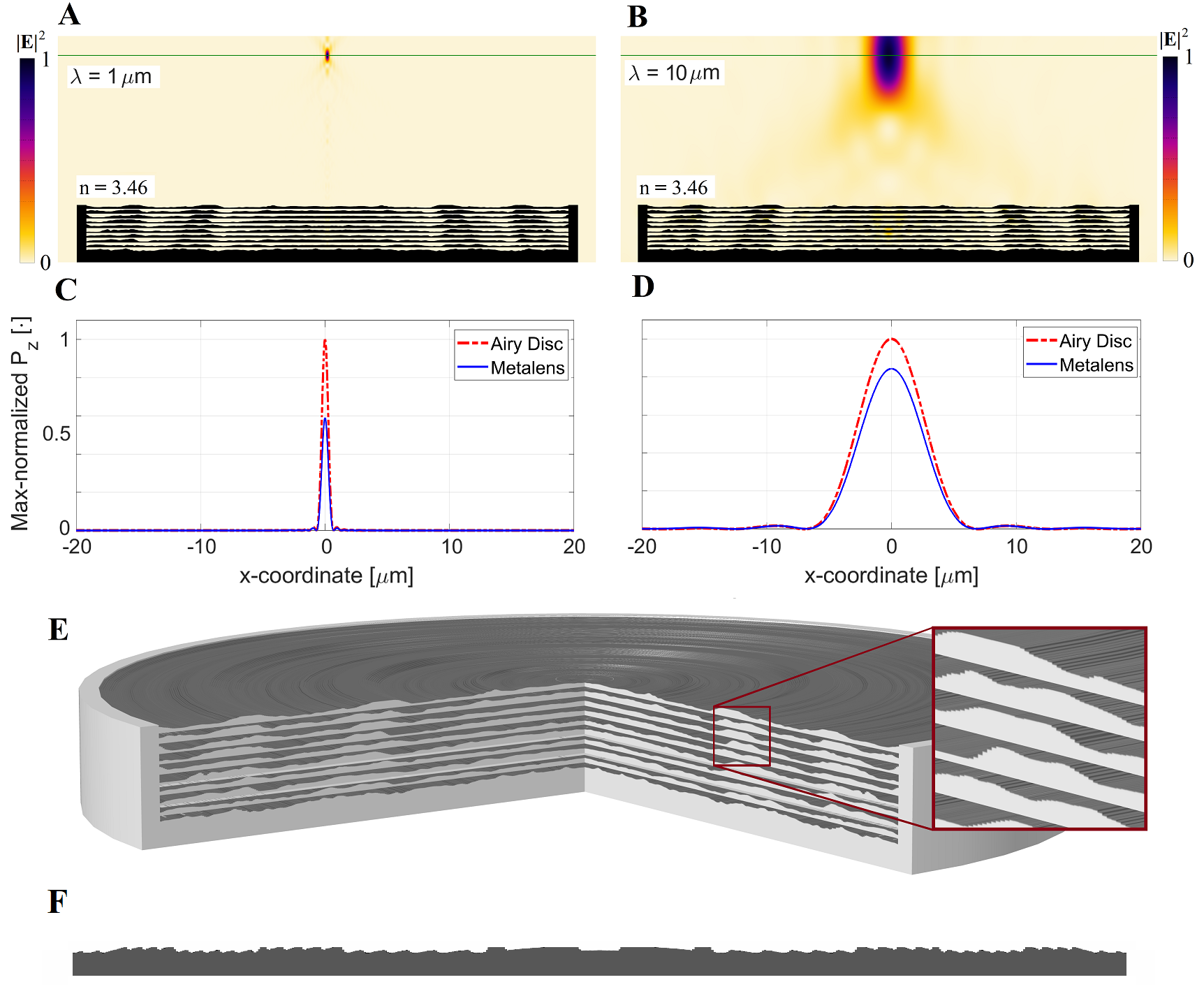}
		\caption{\textbf{A-B)} Max-normalized $|E|^2$-field (thermal colormap) and focal plane (green line) with design overlay (black) in the (x,z)-plane through the center of the lens for \textbf{A)} $\lambda=1 \ \mu$m and \textbf{B)} $\lambda=10000$~nm planewave excitation. \textbf{C-D)} Powerflow in the z-direction through the Focal plane normalized to the maximum of the Airy disc for \textbf{C)} $\lambda=1 \ \mu$m and \textbf{D)} $\lambda=10 \ \mu$m planewave excitation. \textbf{E)} 3D rendering of the metalens design.  {\textbf{F)} Cross section of single-layer reference design.} \label{FIG:LENS_OOMDW}}
	\end{figure}

	As the first example of our framework we tailor a 10-layer silicon ($n=3.46$) in air metalens to focus $\lambda_1 = 1 \ \mu$m light (Fig.~\ref{FIG:LENS_OOMDW}A) and $\lambda_2 = 10 \ \mu$m light (Fig.~\ref{FIG:LENS_OOMDW}B) simultaneously at the same focal spot (NA$=0.85$). The lens is 100~$\mu$m in diameter and has a thickness of 10~$\mu$m. The inverse-designed lens is presented in Fig.~\ref{FIG:LENS_OOMDW}E with the insert showing an example of the layer-height variations. 
	
	From Fig.~\ref{FIG:LENS_OOMDW}A-\ref{FIG:LENS_OOMDW}B it is clear that the lens exhibits the desired numerical aperture (green line). The focusing capability of the lens is found to reach the diffraction-limit for both wavelengths, when measured in terms of the Full Width at Half Maximum (FWHM) of the main lobe in the focal plane (Fig.~\ref{FIG:LENS_OOMDW}C-\ref{FIG:LENS_OOMDW}D). The Strehl ratio (SR) at the two targeted wavelengths, $\lambda_1 = 1 \ \mu$m and $\lambda_2 = 10 \ \mu$m, is computed to be SR $\approx0.60$ and SR $\approx0.84$, respectively. The SR is computed based on the power flow through the focal plane (blue lines) and the corresponding Airy discs (dashed red lines), shown in Fig.~\ref{FIG:LENS_OOMDW}C-\ref{FIG:LENS_OOMDW}D. The absolute power transmission from the substrate of silicon through the lens is computed to be $\mathrm{T}_\mathrm{A,\lambda_1} \approx 82\%$ and $\mathrm{T}_\mathrm{A,\lambda_2} \approx 95\%$, relative to the incident power in the silicon substrate within the lens diameter. Appendix~C includes an additional design example targeting NA$=0.65$ rather than NA=$0.85$ while keeping all other parameters fixed, demonstrating the methods versatility. For that second example we also achieve diffraction-limited focusing and attain Strehl ratios of SR$\approx 0.66$ and SR$\approx 0.99$ for $\lambda_1 = 1 \ \mu$m and $\lambda_2 = 10 \ \mu$m, respectively.
	
	 {To illustrate the benefit of the proposed multi-layer metalens over a single-layer lens we consider a simple single-layer reference design (Fig.~\ref{FIG:LENS_OOMDW}F). The single-layer design is optimized using our proposed approach, with all parameters used in the example held constant, except for the number of layers. For this single-layer reference design we obtain a Strehl Ratio of  $\approx0.37$ and $\approx0.09$ and an absolute transmission efficiency of  $\mathrm{T}_\mathrm{A,\lambda_1} \approx 57\%$ and $\mathrm{T}_\mathrm{A,\lambda_2} \approx 74\%$ at $\lambda_1$ and $\lambda_2$, respectively. Comparing the Strehl ratios and absolute transmission efficiencies to those obtained for the ten layer metalens design, the benefit of multi-layer metalens designs over single-layer designs for this example is clear.}
	
	\section{Tunable multi-wavelength multilayer metalens} \label{SEC:LENS_AM}
	
\begin{figure}[h!]
		\centering\includegraphics[width=1.0\linewidth]{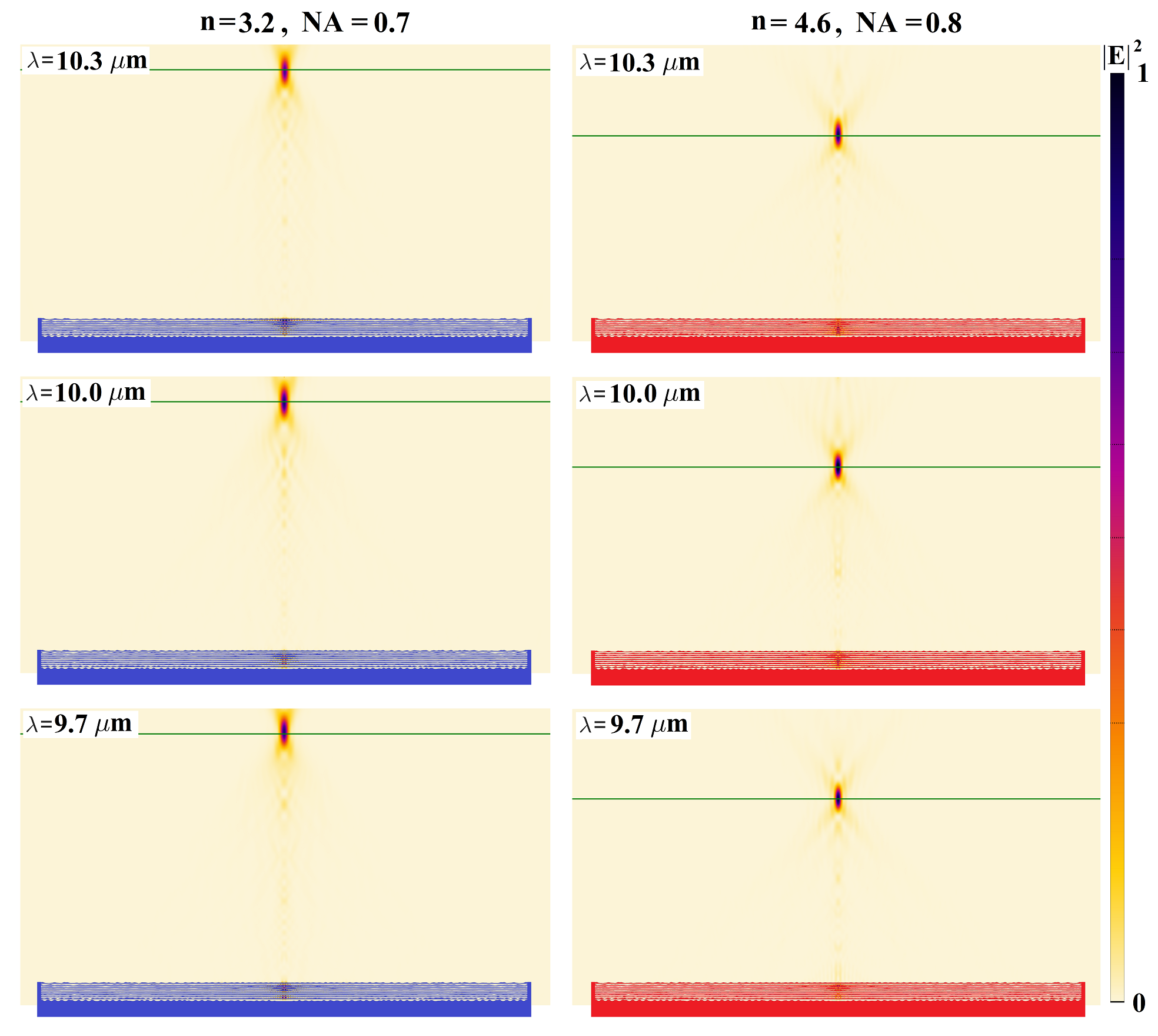}
		\caption{$|E|^2$-field normalized to the largest value across the six cases (thermal colormap) at the [Rows] three targeted wavelengths for the [Columns] two targeted values of the refractive index with the focal plane (green line) and design (black) overlaid. \label{FIG:LENS_FIELD_ILLUSTRATION}}
	\end{figure}

	As a second example of our framework, we design of a 10-layer tunable three-wavelength metalens (see Fig.~\ref{FIG:DESIGN_PROBLEM_ILLUSTRATION}C) capable of shifting the numerical aperture of the lens from NA$=0.7$ (see Fig.~\ref{FIG:LENS_FIELD_ILLUSTRATION}[Left Column]) to NA$=0.8$ (see Fig.~\ref{FIG:LENS_FIELD_ILLUSTRATION}[Right Column]) by changing the refractive index of the active material (GST41T1~\cite{ZHANG_ET_AL_2019}) from $n=3.2$ to $n=4.6 + 0.01 \mathrm{i}$. 
	
	The lens is 625~$\mu$m in diameter and has a thickness of 25~$\mu$m. The lens is designed to operate in the mid-infrared region at wavelengths, $\lambda_1 = 9.7 \ \mu$m (Fig.~\ref{FIG:LENS_FIELD_ILLUSTRATION}[Row 1]), $\lambda_1 = 10 \ \mu$m (Fig.~\ref{FIG:LENS_FIELD_ILLUSTRATION}[Row 2]) and $\lambda_1 = 10.3 \ \mu$m (Fig.~\ref{FIG:LENS_FIELD_ILLUSTRATION}[Row 3]). From Fig.~\ref{FIG:LENS_FIELD_ILLUSTRATION} it is observed that the lens exhibits the desired numerical aperture at all three wavelengths for both values of the refractive index. The Strehl ratio, absolute power transmission and FWHM of the main lobe at the focal point for the three targeted wavelengths and two refractive indices are presented in Tab.~\ref{TAB:TABLE_1}. In brief, a Strehl ratio of approximately $0.5$ is achieved across all six cases with the spatial focusing being at most $11\%$ from the diffraction limit. Finally, a $T_A$ of $\approx 0.3$ for $n = 3.2$ and of $\approx 0.2$ for $n = 4.6 + 0.01 \mathrm{i}$ is achieved.
	
	\begin{table}[]
		\centering
		\begin{tabular}{|l|l|l|l|}
			\hline
			$ \lambda $ & 9.7 $\mu$m  & 10.0 $\mu$m  & 10.3 $\mu$m  \\ \hline
			$n = 3.2$, Strehl ratio $[\cdot]$ & $\approx 0.52$  & $\approx 0.56$  & $\approx 0.55$  \\ \hline
			$n = 4.6 + 0.01 \mathrm{i}$, Strehl ratio $[\cdot]$ & $\approx 0.48$  & $\approx 0.54$  & $\approx 0.55$  \\ \hline
			$n = 3.2$, FWHM main lobe $\left[\frac{\lambda}{2\mathrm{NA}}\right]$ & $\approx 1.11$  & $\approx 1.08$ & $\approx 1.08$ \\ \hline
			$n = 4.6 + 0.01 \mathrm{i}$, FWHM main lobe $\left[\frac{\lambda}{2\mathrm{NA}}\right]$ & $\approx 1.00$ & $\approx 1.07$ & $\approx 1.00$ \\ \hline
			$n = 3.2$, $T_A$ $\left[\frac{\textbf{P}_{\mathrm{lens}}}{\textbf{P}_{\mathrm{inc}}}\right]$ & $\approx 0.31$  & $\approx 0.33$ & $\approx 0.29$ \\ \hline
			$n = 4.6 + 0.01 \mathrm{i}$, $T_A$ $\left[\frac{\textbf{P}_{\mathrm{lens}}}{\textbf{P}_{\mathrm{inc}}}\right]$ & $\approx 0.22$ & $\approx 0.23$ & $\approx 0.20$ \\ \hline
		\end{tabular}
		\caption{Strehl ratio, FWHM of main lobe in the focal plane and the absolute power transmission relative to the incident power in the Si substrate for the lens in Fig.~\ref{FIG:LENS_FIELD_ILLUSTRATION}. \label{TAB:TABLE_1}}
	\end{table}
	
	\section{Experimental validation of single-layer variable-height metalens} \label{SEC:SINGLE_LAYER_EXPERIMENTAL_VALIDATION}
	
	Finally, as a proof of concept, we demonstrate experimentally that the proposed method can be used to design variable-height metasurfaces for given fabrication specifications (details about the fabrication and experiment are given in Appendix~D and Appendix~E). Figure~\ref{FIG:SINGLE_LAYER_EXPERIMENTAL_VALIDATION}\textbf{G} shows a 3D rendering of the designed single-layer varying-height metalens. The metalens is fabricated via 3D two-photon polymerization in IP-Dip, a low-refractive-index polymer \cite{gissibl2017refractive} that can be printed in voxel sizes with in-plane feature sizes $\sim 100$ nm and fixed voxel aspect ratio of $\sim 1$ to 3. This example is not aimed at designing the largest area lens possible nor at achieving the highest possible numerical performance, but at designing a metalens that complies with fabrication constraints. In this respect, the design is restricted to a diameter of 200 $\mu$m with a 300~nm radial pixel size and a varying height with a maximum height of 900~nm, restricted to height-variations in 100~nm increments. The height of the individual radial pixel is allowed to vary independently of its neighbors (i.e. no filtering is applied to $\xi_{L}$). 
	
	The lens is designed to focus $\lambda = 1550$ nm light at normal incidence with a numerical aperture of $0.4$. The numerically computed electric-field intensity at 1550 nm for planewave illumination of the lens at normal incidence is shown in Fig.~\ref{FIG:SINGLE_LAYER_EXPERIMENTAL_VALIDATION}A, clearly showing that the targeted numerical aperture (green line) is achieved. Numerically the lens achieves near diffraction-limited focusing in terms of the FWHM of main lobe of the power flow in the z-direction through the focal plane. A FWHM of $\approx1000$~nm is computed numerically, corresponding to $\approx3.2\%$ above the diffraction limit (Using the theoretical limit $\frac{\lambda}{2\mathrm{NA}}\approx969$ nm). 
	
	The absolute power transmission from the substrate of IP-Dip through the lens is computed at $\mathrm{T}_\mathrm{A} \approx 93\%$, relative to the incident power in the IP-Dip substrate within the lens diameter. A Strehl ratio of $\mathrm{SR} \approx 0.29$ is computed by numerical integration of the power flow over the focal plane. This SR value reveals that a significant fraction of the power is not flowing through the focal point. From a design point of view, the Strehl ratio is easy to improve using our framework by increasing the design freedom, either by changing the metasurface material; by decreasing the radial pixel size; by increasing the number of height increments; by increasing the total height of the lens and/or by introducing multiple-layers in the lens. All of these were demonstrated in the two previous examples.
	
	\begin{figure}[h!]
		\centering\includegraphics[width=1.0\linewidth]{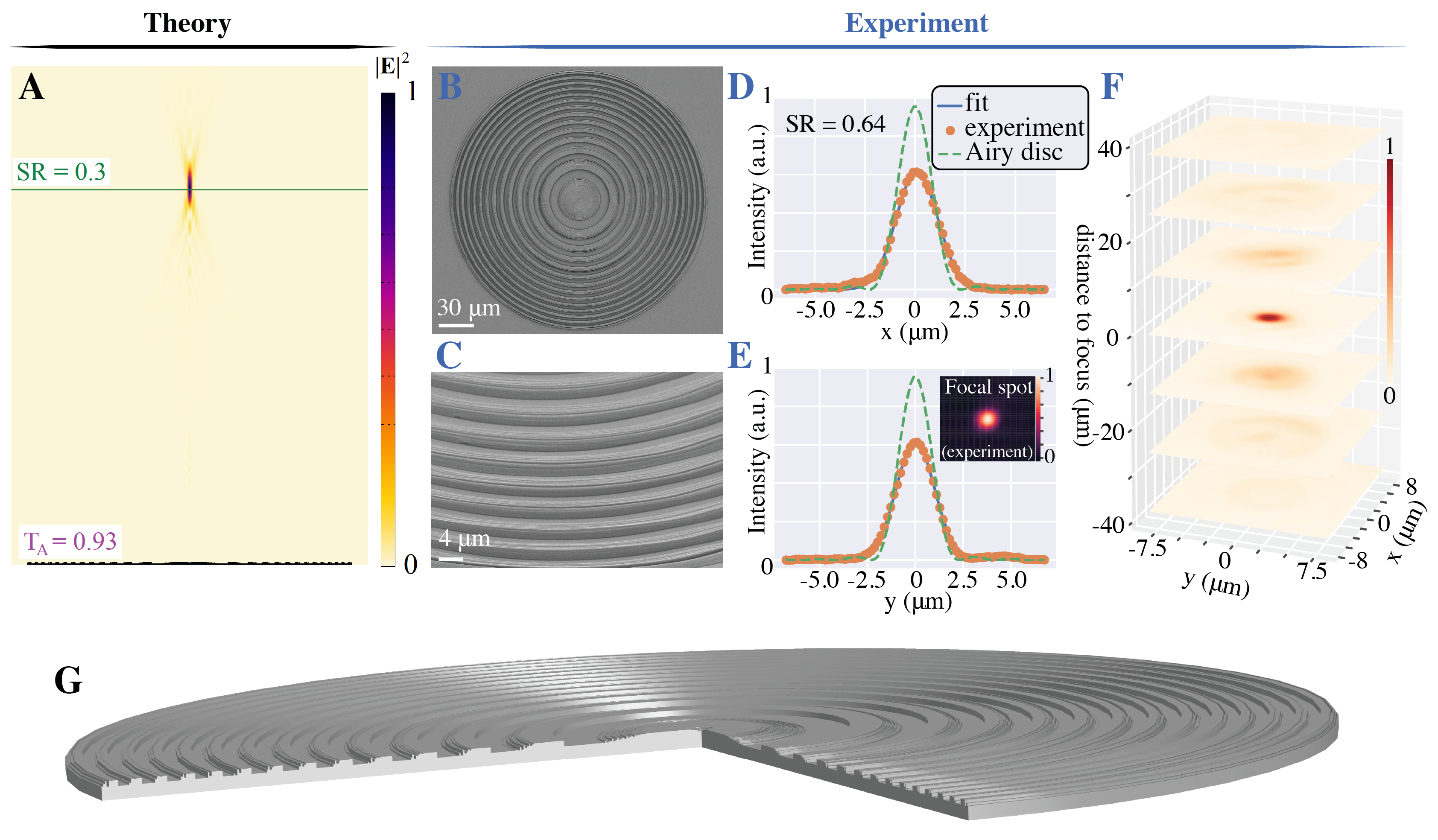}
		\caption{3D-printed single-layer circular symmetric metasurface. \textbf{A)} Max-normalized electric field intensity $\vert \textbf{E} \vert^2$ (thermal colormap) and focal plane (green line) with design overlay (black). \textbf{B)} Scanning Electron Micrograph of the full lens. Scale bar = 30 $\mu$m. \textbf{C)} Scanning Electron Micrograph of a smaller area, showing the height variation along the radial direction. Scale bar = 4 $\mu$m. \textbf{D)} Horizontal cut of the focal spot, showing a Gaussian fit to the spot and the corresponding Airy disk (which defines the Strehl Ratio). \textbf{E)} Vertical cut of the focal spot. The inset shows the focal spot recorded by the imaging setup (measured on the NIR imaging camera). \textbf{F)} Focal spot measured at various positions along the optic axis. \textbf{G)} 3D rendering of metalens design.
			\label{FIG:SINGLE_LAYER_EXPERIMENTAL_VALIDATION}}
	\end{figure}
	
	Experimentally the Strehl ratio is estimated to be $\approx 0.64$ by integrating the power flow over an $8~\mu\mathrm{m} \times 8.5~\mu\mathrm{m}$ region centered at the focal spot. Computing the SR numerically using the same integration area we obtain $\mathrm{SR} \approx 1.0$ showing that a majority of the power transmitted through the lens is not focused at the focal spot but flows through the focal plane outside this area. The discrepancy between the experimentally measured and numerically computed SR suggests that the experiment overestimates the SR, due to the camera's limited field of view. This is supported by the relatively low measured absolute focusing efficiency of $\approx 5 \%$. The measured focal spot (Fig.~\ref{FIG:SINGLE_LAYER_EXPERIMENTAL_VALIDATION}D-\ref{FIG:SINGLE_LAYER_EXPERIMENTAL_VALIDATION}F) exhibits FWHMs of $2.28 \pm 0.16 ~\mu m$ (resp. $2.22\pm0.17~\mu m$) along the horizontal (resp. vertical) direction, corresponding to $18 \pm 8 \%$ (resp. $15 \pm 9\% $) above the diffraction limit. These experimental results validate the feasability of freeform axisymmetric metasurfaces experimentally. While this proof-of-concept experiment was limited to a single-layer metasurface, the radially-varying height of the structure can, to the authors knowledge, only be implemented with fabrication techniques such as 2.5D lithography or multi-photon polymerization. This is a first step towards realizing the full potential of the freeform axisymmetric inverse design technique presented in this work.  
	
	 {Achieving true multi-layer closely-packed metasurfaces presents additional challenges, such as the accurate positioning and alignment of each layer. Yet another challenge -- which is specific to two-photon polymerization -- is to design structures that allow unpolymerized material to be extracted, a constraint that could be included in further refinements of our theory.}
	
	\section{Conclusion}  \label{SEC:CONCLUSION}
	
	In this paper, we demonstrated that fullwave Maxwell Equation based inverse design of axisymmetric structures can tackle challenging new design problems involving radically different wavelengths or active materials. We believe that the proposed design framework opens the way to many new applications whose functionality goes far beyond traditional lenses, such as end-to-end design~\cite{lin2020end}, hyperspectral imaging~\cite{kristina2020spectral}, depth sensing~\cite{guo2019compact} and nonlinear imaging~\cite{schlickriede2020nonlinear}.  {While we expect a small-angle paraxial regime to be valid for our lens designs, which may be used for imaging over a narrow field of view, we will consider, in a future work, thorough corrections of off-axis as well as chromatic aberrations in a single-piece axisymmetric metalens design.}
	
	 {An example of the significant performance benefits that can be attained by designing multi-layer metalenses, compared to single-layer metalenses, was given in Sec.~\ref{SEC:LENS_OOMDW}. The relationship between the targeted number of layers and the performance of the lens has not been investigated in detail and such a study for different metalens applications is likely to provide valuable information and is thus interesting to pursue.}
	
	Computationally, there are  {several} ways to scale our algorithm to much larger designs. The simplest would be to  {utilize} near-to-farfield transformations~\cite{taflove2005computational} to omit  {simulation of} the homogeneous region above the lens from the computation, which would allow us to increase the radial size by a factor of $\sim 10$. Approximate domain-decomposition could be used to partition a larger lens into overlapping subdomains solved in parallel (but optimized together)~\cite{lin2019overlapping}. To increase design freedom, the axisymmetry could be relaxed to various forms of $N$-fold or other rotational symmetries. One could also explore fully free-form topology optimization for 3D-printed structures with manufacturability constraints~\cite{CHRISTIANSEN_SMO_2015,ZHOU_SMO_2015,WANG_ET_AL_2018,Li2016}.
	
	
	 {When employing the proposed approach for materials with a large non-zero extinction coefficient, $\kappa$, i.e. a complex refractive index $\tilde{n} = n + \mathrm{i} \kappa$, it is possible that one needs to consider a different material interpolation scheme, to achieve high quality results from the inverse design process \cite{CHRISTIANSEN_2019}.}
	
	Experimentally, we have shown a proof-of-concept fabricated structure using a two-photon 3D-lithography process. The inverse-designed metasurface achieved focusing at the telecommunication wavelength of 1550 nm, close to the diffraction limit, with a numerical aperture of 0.4. In the future, we will develop multilayer fabrication of these structures, in order to realize the full potential of the design technique developed in this work. A key challenge is to realize mechanically stable multilayered structures from which unpolymerized resist can be extracted. Application-specific two-photon polymerization setups \cite{fang2020multilevel} can achieve more height levels and some control over the voxel aspect ratio.  For devices operating at shorter wavelengths, thus requiring proportionately smaller feature sizes, the design process would shift to multilayer structures with piecewise-constant cross-section~\cite{lin2019topology, piggott2017fabrication}.  Conversely, at longer wavelengths such as for microwave wavefront shaping, multilayer structures could be straightforwardly fabricated, for instance, by stacking multiple stacks of 3D-printed resins or drilled materials \cite{camayd2020multifunctional}.
	
\section*{Funding} 
This work was supported in part by Villum Fonden through the NATEC (NAnophotonics for TErabit Communications) Centre (grant no.~8692); the Danish National Research Foundation through NanoPhoton Center for Nanophotonics (grant no.~DNRF147); the U.~S.~Army Research Office through the Institute for Soldier Nanotechnologies (award no.~W911NF-18-2-0048); and the MIT-IBM Watson AI Laboratory (challenge no.~2415).

\section*{Acknowledgements} 

 Y.S. acknowledges the Swiss National Science Foundation (SNSF) through Project No. P2EZP2\textunderscore188091.

\section*{Disclosures} 

The authors declare that there are no conflicts of interest related to this article.

	\bibliographystyle{ieeetr} 
	\bibliography{References_arXiv}
	
	\appendix
	\section*{Appendices}
	
	\section*{Appendix A. Optimization and Numerical Modelling}  \label{APN:OPTIMIZATION_EVALUATION}
	
	The physics is modelled in COMSOL Multiphysics \cite{COMSOL55} and the optimization problem is solved using the Globally Convergent Method of Moving Asymptotes (GCMMA) \cite{SVANBERG_2002}.  \\
	
	\noindent In the design process $\Omega_\mathrm{D}$ and the solid material regions in $\Omega$ are discretized using a structured quadrilateral mesh, while the surrounding air regions are discretized using an unstructured triangular mesh, both of which uses $\geq 10$ elements per $\lambda / n$. The finite element method with a linear Lagrangian basis is used to discretize the physics \cite{BOOK_FEM_EM_JIN}. 
	
	\noindent The following stopping criterion is used to terminate the iterative solution of the optimization problem: \\
	
	\begin{algorithmic}
		\If {$i \geq i_{\mathrm{min}}$}
		\If {$ \vert \Phi_{i} - \Phi_{i-n} \vert / \vert \Phi_{i}  \vert  \leq 0.01 \ \forall \ n  \  \lbrace  1,2,...,10 \rbrace $}  
		\State Terminate optimization.
		\EndIf
		\EndIf \\
	\end{algorithmic}
	
	\noindent Here $i$ denotes the current optimization iteration, $i_{\mathrm{min}} = 70$ denotes the minimum number of design iterations taken. $\Phi_{i}$ denotes the objective function value at the $i$'th iteration and $n \in \mathbb{N}^{+}$. \\
	
	\section*{Appendix B. Study parameters} \label{APN:STUDY_PARAMETERS}
	
	The parameters used in setting up the models and associated optimization problems for the three examples follow here.
	
	\subsection*{B. 1. Multi-scale multi-wavelength multilayer metalens}
	
	For the problem treated in Sec.~\ref{SEC:LENS_OOMDW} the following parameter values are used: \\
	
	\noindent The axisymmetric model domain $\boldmath{\Omega}$ has a width of 57~$\mu$m in the r-direction and a height of 82~$\mu$m in the z-direction. $\boldmath{\Omega}$ is surrounded on three of four sides by a perfectly matched layer with a depth of 1500~nm (Fig.~\ref{FIG:DESIGN_PROBLEM_ILLUSTRATION}). The metalens design domain $\boldmath{\Omega}_{\mathrm{D}}$ is taken to have a radius of 50~$\mu$m and a height of 10~$\mu$m and is separated into ten layers of equal height. Each layer has a total height of 1~$\mu$m with the designable region having a height of 600~nm and the fixed air and silicon regions each having heights of 200 nm. It is placed on a slab of material of 2~$\mu$m thickness placed at the bottom edge of the model domain. \\
	
	\noindent The radial design pixel size is restricted to a minimum of 200~nm and the height-variation is restricted to 25~nm increments. \\
	
	\noindent The two wavelengths of the incident field are taken to be $\lambda_1 = 1$~$\mu$m and $\lambda_2 = 10$~$\mu$m. The lens is taken to be made of silicon in an air background. The refractive index of air are taken to be $n_{\mathrm{air}} = 1.0$. The refractive index of silicon is taken to be $n_{\mathrm{si}} = 3.46$ at both operating wavelengths. The speed of light is taken to be $c = 3 \cdot 10^8$ m/s. The numerical aperture is taken to be NA$= 0.65$. \\
	
	\noindent The initial guess for the design field is $\xi_{L,\mathrm{initial}}(\textbf{r}) = 0.5 \ \forall \ \textbf{r} \in \boldmath{\Omega}_{\mathrm{D}}$ for all 10 layers. A filter radius of $r_f = 400$~nm is used to limit the gradient of the heigh variation in each layer to avoid rapid pixel-by-pixel oscillations in the design. The value of the thresholding sharpness parameters is $\beta = 40$.

	\subsection*{B. 2. Tunable multi-wavelength multilayer metalens}
	
	For the problem treated in Sec.~\ref{SEC:LENS_AM} the following parameter values are used: \\
	
	\noindent The axisymmetric model domain $\boldmath{\Omega}$ has a width of 342.5~$\mu$m in the r-direction and a height of 380~$\mu$m in the z-direction. $\boldmath{\Omega}$ is surrounded on three of four sides by a perfectly matched layer with a depth of 15~$\mu$m (Fig.~\ref{FIG:DESIGN_PROBLEM_ILLUSTRATION}). The metalens design domain $\boldmath{\Omega}_{\mathrm{D}}$ is taken to have a radius of 312.5~$\mu$m and a height of 25~$\mu$m and is separated into ten layers of equal height with a 2000~nm designable region and 250~nm fixed air region and 250~nm fixed solid region. It is placed on a slab of material of 5~$\mu$m thickness placed at the bottom edge of the model domain. \\
	
	\noindent The radial design pixel size is restricted to a minimum of 600~nm and the height-variation is restricted to 100~nm increments. \\
	
	\noindent The three wavelengths of the incident field are taken to be $\lambda_1 = 9.7$~$\mu$m, $\lambda_1 = 10$~$\mu$m and $\lambda_2 = 10.3$~$\mu$m. The lens is taken to be made of GST41T1 in an air background. The refractive index of air are taken to be $n_{\mathrm{air}} = 1.0$. The refractive index of the active material is taken to be $n_{\mathrm{GST,1}} = 3.2$ in the first configuration and $n_{\mathrm{GST,2}} = 4.6$ in the second at all operating wavelengths. The speed of light is taken to be $c = 3 \cdot 10^8$ m/s. The numerical aperture of the lens is taken to be NA$= 0.7$ in the first configuration and NA$=0.8$ in the second. \\
	
	\noindent The initial guess for the design field is $\xi_{L,\mathrm{initial}}(\textbf{r}) = 0.5 \ \forall \ \textbf{r} \in \boldmath{\Omega}_{\mathrm{D}}$ for all 10 layers. A filter radius of $r_f = 3~\mu$m is used to limit the gradient of the height-variation in each layer (see the insert in Fig.~\ref{FIG:DESIGN_PROBLEM_ILLUSTRATION}\textbf{C}). The value of the thresholding sharpness parameters is $\beta = 40$.
	
	\subsection*{B. 3. Single-layer variable-height metalens}
	
	For the problem treated in Sec.~\ref{SEC:SINGLE_LAYER_EXPERIMENTAL_VALIDATION} the following parameter values are used: \\
	
	\noindent The axisymmetric model domain $\boldmath{\Omega}$ has a width of 106~$\mu$m in the r-direction and a height of 301.8~$\mu$m in the z-direction. $\boldmath{\Omega}$ is surrounded on three of four sides by a perfectly matched layer with a depth of 3~$\mu$m (Fig.~\ref{FIG:DESIGN_PROBLEM_ILLUSTRATION}). The metalens design domain $\boldmath{\Omega}_{\mathrm{D}}$ is taken to have a radius of 100~$\mu$m and a height of 900~nm and comprises a single layer constituting the designable region. The design domain is placed on a slab of material of 500~nm thickness placed at the bottom edge of the model domain. \\
	
	\noindent The design is discretized into 300~nm radial increments and 100~nm height increments. \\
	
	\noindent The wavelength of the incident field is taken to be $\lambda = 1550$~nm. The lens is taken to be made of IP-Dip in an air background. The refractive index of air are taken to be $n_{\mathrm{air}} = 1.0$. The refractive index of IP-Dip is taken to be $n_{\mathrm{si}} = 1.507$ at both operating wavelengths. The speed of light is taken to be $c = 3 \cdot 10^8$ m/s. The numerical aperture is taken to be NA$= 0.4$. \\
	
	\noindent The initial guess for the design field is $\xi_{L,\mathrm{initial}}(\textbf{r}) = 0.5 \ \forall \ \textbf{r} \in \boldmath{\Omega}_{\mathrm{D}}$. No smoothing filter is applied. The value of the thresholding sharpness parameters is $\beta = 40$.

	\section*{Appendix C. Second example of a multi-scale multi-wavelength multilayer metalens design} \label{APN:ADDITIONAL_OOMDW_EXAMPLE}
	
\begin{figure}[h!]
		\centering\includegraphics[width=1.0\linewidth]{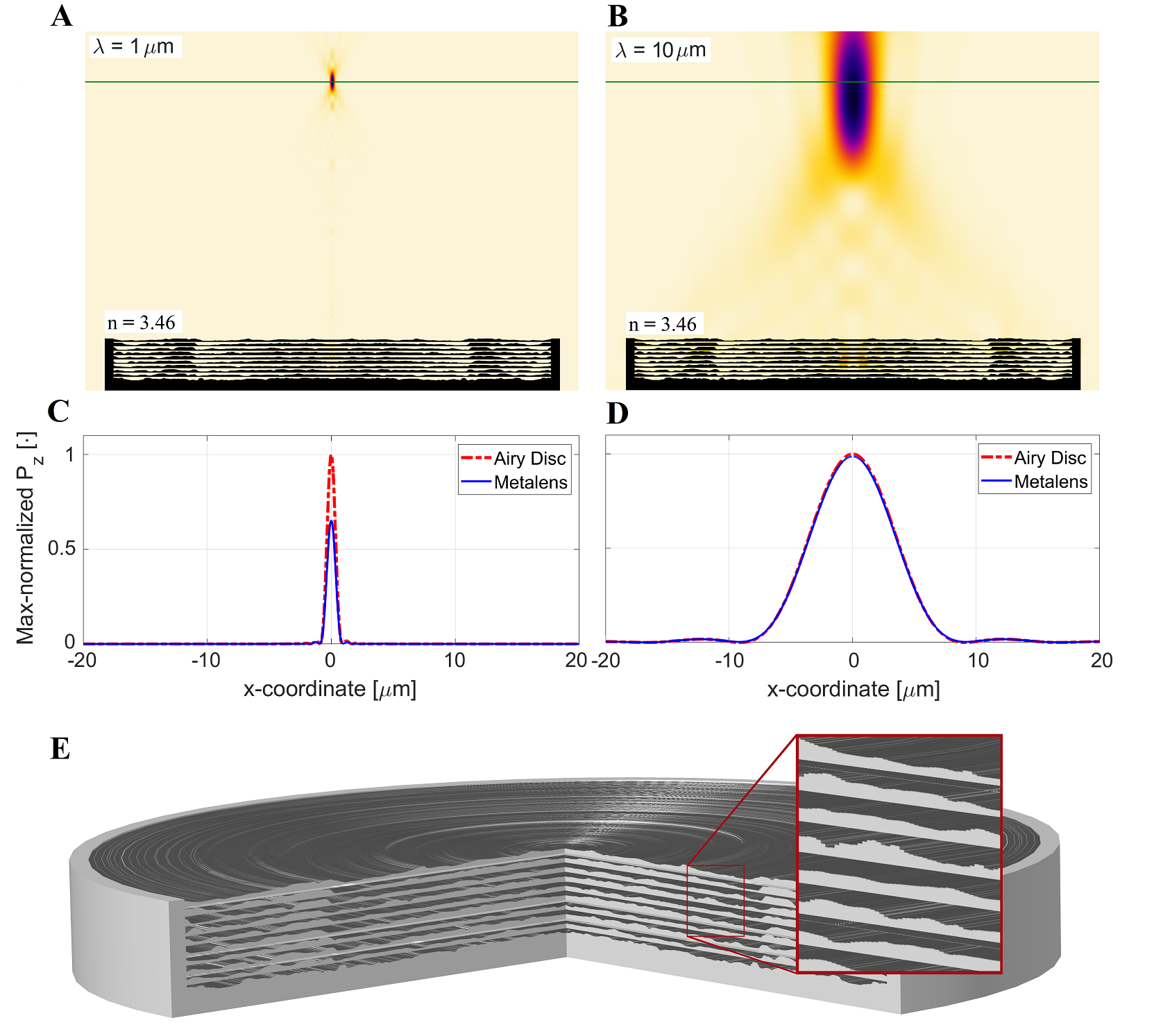}
		\caption{\textbf{A-B)} Max-normalized $|E|^2$-field (thermal) and focal plane (green line) with design overlay (black) in the (x,z)-plane through the center of the lens for \textbf{A)} $\lambda=1 \ \mu$m and \textbf{B)} $\lambda=10000$~nm planewave excitation. \textbf{C-D)} Powerflow in the z-direction through the Focal plane normalized to the maximum of the Airy disc for \textbf{C)} $\lambda=1 \ \mu$m and \textbf{D)} $\lambda=10 \ \mu$m planewave excitation. \textbf{E)} 3D rendering of the metalens design. \label{FIG:LENS_OOMDW_SECOND_EXAMPLE}}
	\end{figure}

	We tailor a 10-layer silicon ($n=3.46$) in air metalens to focus $\lambda_1 = 1 \ \mu$m light (Fig.~\ref{FIG:LENS_OOMDW}\textbf{A)} and $\lambda_2 = 10 \ \mu$m light (Fig.~\ref{FIG:LENS_OOMDW}\textbf{B}) simultaneously at the same focal spot (NA$=0.65$). The lens has identical dimensions and design resolution as the lens in Sec.~\ref{SEC:LENS_OOMDW}. The final lens design is presented in Fig.~\ref{FIG:LENS_OOMDW_SECOND_EXAMPLE}\textbf{E} with the insert showing an example of the layer-height variations. 
	
	Figures~\ref{FIG:LENS_OOMDW}\textbf{A}-\ref{FIG:LENS_OOMDW}\textbf{B} show that the lens exhibits the desired numerical aperture at both wavelengths (green line). Further, the focusing capability of the lens is diffraction-limited for both wavelengths. The Strehl ratio (SR) at the two targeted wavelengths, $\lambda_1 = 1 \ \mu$m and $\lambda_2 = 10 \ \mu$m, is computed to SR $=\approx0.66$ and SR $=\approx0.99$, respectively, from the data in Fig.~\ref{FIG:LENS_OOMDW}\textbf{C}-\ref{FIG:LENS_OOMDW}\textbf{D}. 
	
	\section*{Appendix D. Fabrication}
	The metalens was fabricated using a commercial two-photon polymerization system (Nanoscribe Photonic Professional GT) on a $700$-micron-thick fused silica substrate, where the structures are written in circles with height increments of $~ 100 nm$. For this purpose, piezo actuators move the sample in the out-of-plane direction after fabricating each layer. Geometrical parameters and dose (scanning speed and laser power) are optimized with a dose test on this specific machine. In the in-plane direction the laser beam is guided by galvanometric mirrors parallel to the substrate. After printing, the structures are put in a developer bath (PGMEA 5~min) and dried in IPA with a critical point dryer Auto Samdri 815 Series A.
	
	\section*{Appendix E. Experiment}
	
\begin{figure}
		\centering
		\includegraphics[width=1.0\linewidth]{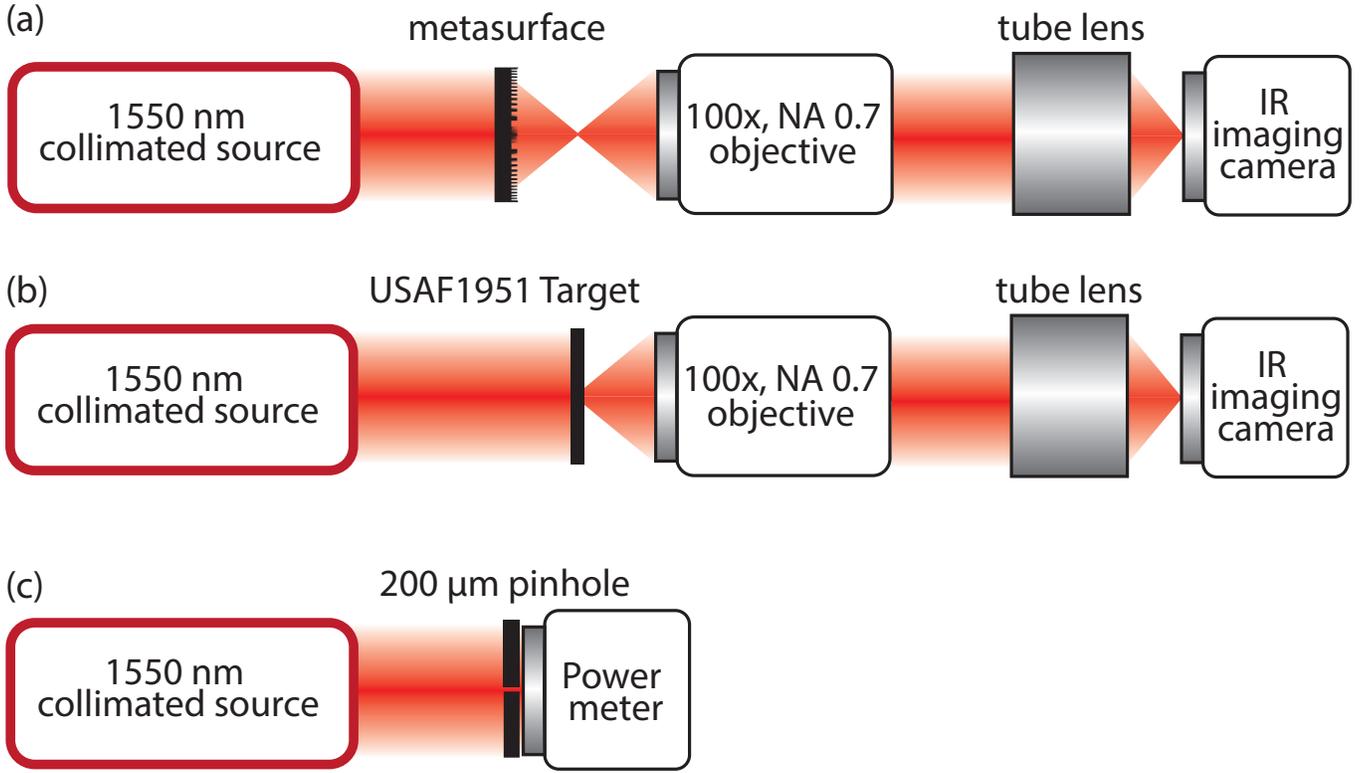}
		\caption{Experimental setup. (a) Experimental configuration to measure the metasurface performance (focal spot, cross-sections, efficiency, Strehl Ratio). The objective-tube lens-camera assembly can be translated along the optic axis. The setup is first calibrated by imaging a USAF1951 target (b). To calibrate our power estimates, we measured the power going through a pinhole with the setup shown in (c).}
		\label{FIG:EXPT_SETUP}
	\end{figure}

	For the proof-of-concept experimental results presented in Fig.~ \ref{FIG:SINGLE_LAYER_EXPERIMENTAL_VALIDATION}, we used the imaging setup shown in Fig.~\ref{FIG:EXPT_SETUP}(a). A Ando AQ4321D Tunable Laser Source produces a fiber-coupled output at 1550~nm. The fiber output is collimated with a set of lenses. In the measuring configuration Fig.~\ref{FIG:EXPT_SETUP}, the collimated beam is focused by the metasurface, and the focal spot is imaged by an objective - tube lens - IR imaging camera system. For this measurement, we used a 100X Mitutoyo Plan Apo NIR HR Infinity Corrected Objective, a ThorLabs $f = 200$mm tube lens, and a EC MicronViewer 7290A. The imaging setup was first calibrated using the configuration shown in Fig.~\ref{FIG:EXPT_SETUP}(b), where the equivalent pixel size on the detector is evaluated by imaging a USAF1951 target. To evaluate the efficiency of the metasurface, we measured the equivalent power going through a $200~\mu$m diameter pinhole with the configuration shown in Fig.~\ref{FIG:EXPT_SETUP}(c).
	
	To estimate the metasurface efficiency, we use the intensity-voltage relation of the NIR camera provided by the vendor. It has the form $I = K V_s^{1/g}$, where $I$ is the  incident optical power on a pixel, $V_s$ the generated voltage at that pixel, and $g$ the characteristic nonlinear slope of the intensity-voltage relation, which is given to be $g \sim 0.7$. We first calibrate the proportionality constant $K$ by measuring the signal produced by the camera of a known beam power. This allows us to translate the measured voltage on a pixel to an incident power (in W). We also measure the incident intensity on the metasurface area with the experimental configuration shown in Fig.~\ref{FIG:EXPT_SETUP}(c).
	The efficiency is then calculated as
	\begin{equation*}
	\text{Eff} = \frac{K}{L} \frac{\sum_{i \in \text{pixels in focal spot}} V_i^{1/g}  }{P_\text{ref}},
	\end{equation*}
	where $L$ is the estimated optical loss through the objective and tube lens, which is 0.55 (objective) $\times$ 0.88 (tube lens).
	We typically remove the background from the measured focal spot in order to estimate the metasurface efficiency.

\end{document}